\documentclass{SciPost}

\hypersetup{
    colorlinks,
    linkcolor={red!50!black},
    citecolor={blue!50!black},
    urlcolor={blue!80!black}
}

\usepackage[bitstream-charter]{mathdesign}
\DeclareSymbolFont{usualmathcal}{OMS}{cmsy}{m}{n}
\DeclareSymbolFontAlphabet{\mathcal}{usualmathcal}

\fancypagestyle{SPstyle}{
\fancyhf{}
\lhead{\raisebox{-1.5mm}[0pt][0pt]{\href{https://scipost.org}{\includegraphics[width=20mm]{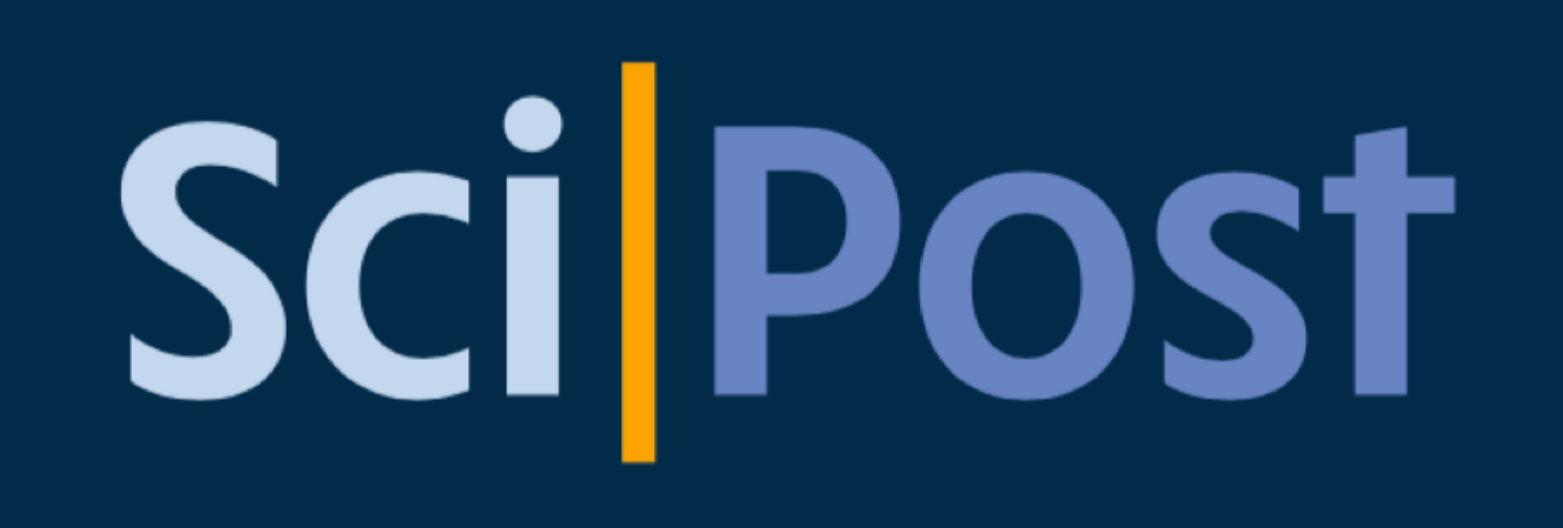}}}}
\rhead{\small \href{https://scipost.org/SciPostPhysProc.4.002}{SciPost Phys. Proc. 4, 002 (2021)}}

\fancyfoot[C]{\textbf{002.\thepage}}
}

 \def\be{\begin{equation}}
 \def\ee{\end{equation}}

 \def\A{\mathcal{A}}

 \def\2{\frac{1}{2}}
 \def\4{\frac{1}{4}}

\begin{document}

\pagestyle{SPstyle}

\begin{center}{\Large \textbf{\color{scipostdeepblue}{
De Sitter entropy as holographic entanglement entropy\\
}}}\end{center}

\begin{center}
\textbf{
Nikolaos Tetradis\textsuperscript{$\star$}
}
\end{center}

\begin{center}
Department of Physics,
National and Kapodistrian University of Athens,\\ University Campus, Zographou 157 84, Greece
\\[\baselineskip]
$\star$ \href{mailto:ntetrad@phys.uoa.gr}{\small \sf ntetrad@phys.uoa.gr}
\end{center}

\definecolor{palegray}{gray}{0.95}
\begin{center}
\colorbox{palegray}{
  \begin{tabular}{rr}
  \begin{minipage}{0.1\textwidth}
  \end{minipage}
  &
  \begin{minipage}{0.75\textwidth}
    \begin{center}
    {\it 4th International Conference on Holography,\\ 
    String Theory and Discrete Approach}\\
    {\it Hanoi, Vietnam, 2020} \\
    \doi{10.21468/SciPostPhysProc.4}\\
    \end{center}
  \end{minipage}
\end{tabular}
}
\end{center}

\section*{\color{scipostdeepblue}{Abstract}}
{\bf
We review the results of refs. \cite{tetradis1,tetradis2}, in which 
the entanglement entropy in spaces with horizons, such as Rindler or de
Sitter
space, is computed using holography.  This is achieved through an appropriate
slicing of anti-de Sitter space and the implementation of a UV cutoff.
When the entangling surface coincides with the horizon of the boundary
metric,
the entanglement entropy can be identified with the standard gravitational
entropy of the space. For this to hold, the effective Newton's constant must
be defined appropriately by absorbing the UV cutoff.  Conversely, the UV
cutoff
can be expressed in terms of the effective Planck mass and the number of
degrees of freedom of the dual theory. For de Sitter space, the entropy is
equal to the Wald entropy for an effective action that includes the
higher-curvature terms associated with the conformal anomaly. The
entanglement
entropy takes the expected form of the de Sitter entropy, including
logarithmic corrections.
}

\begin{center}
\begin{tabular}{lr}
\begin{minipage}{0.56\textwidth}
{\small Copyright N. Tetradis. \newline
This work is licensed under the Creative Commons \newline
\href{http://creativecommons.org/licenses/by/4.0/}{Attribution 4.0 International License}. \newline
Published by the SciPost Foundation.
}
\end{minipage}
&
\begin{minipage}{0.44\textwidth}
\noindent\begin{minipage}{0.68\textwidth}
{\small Received 22-10-2020 \newline Accepted 09-11-2020 \newline Published 13-08-2021
}
\end{minipage}
\\\\
\small{\doi{10.21468/SciPostPhysProc.4.002}
}
\end{minipage}
\end{tabular}
\end{center}


\noindent\rule{\textwidth}{1pt}
\vspace{10pt}


The fact that the divergent part of the entanglement entropy
scales with the area of the entangling surface \cite{sorkin,sorkin2}  
suggests a connection with the gravitational entropy of spaces containing   
horizons. It seems reasonable that the entropies should become equal when 
the entangling surface is identified with a horizon.
We address this problem in the context of the AdS/CFT correspondence
through use of appropriate coordinates that set 
the boundary metric in Rindler or static
de Sitter form.
According to the Ryu-Takayanagi proposal \cite{ryu,ryu2,ryu3}, the 
entanglement entropy of a part of the AdS boundary within
an entangling surface $\A$ is proportional to the
area of a minimal surface 
$\gamma_A$ anchored on $\A$ and extending into the bulk. 

We consider the standard parameterization of $(d+2)$-dimensional AdS space 
with global coordinates, as well as parametrizations through 
Fefferman-Graham coordinates, with the boundary located at the value
$z=0$ of the bulk coordinate.
As a first case we consider a metric with a Rindler boundary:
\be
ds^2_{d+2}
= \frac{R^2}{z^2}\left[dz^2
-a^2 y^2 d\eta^2+dy^2 + d \vec{x}_{d-1} \right],
\label{rindler} \ee
where $a$ is a constant parameter. The timelike coordinate $\eta$ takes values
$-\infty < \eta < \infty$. The range  
$0< y<\infty$ of the spacelike coordinate $y$ covers the right (R) Rindler wedge, while
the range $-\infty < y <0$ covers the 
left (L) wedge.

\begin{figure}[t]
\centering
$$
\includegraphics[width=0.42\textwidth]{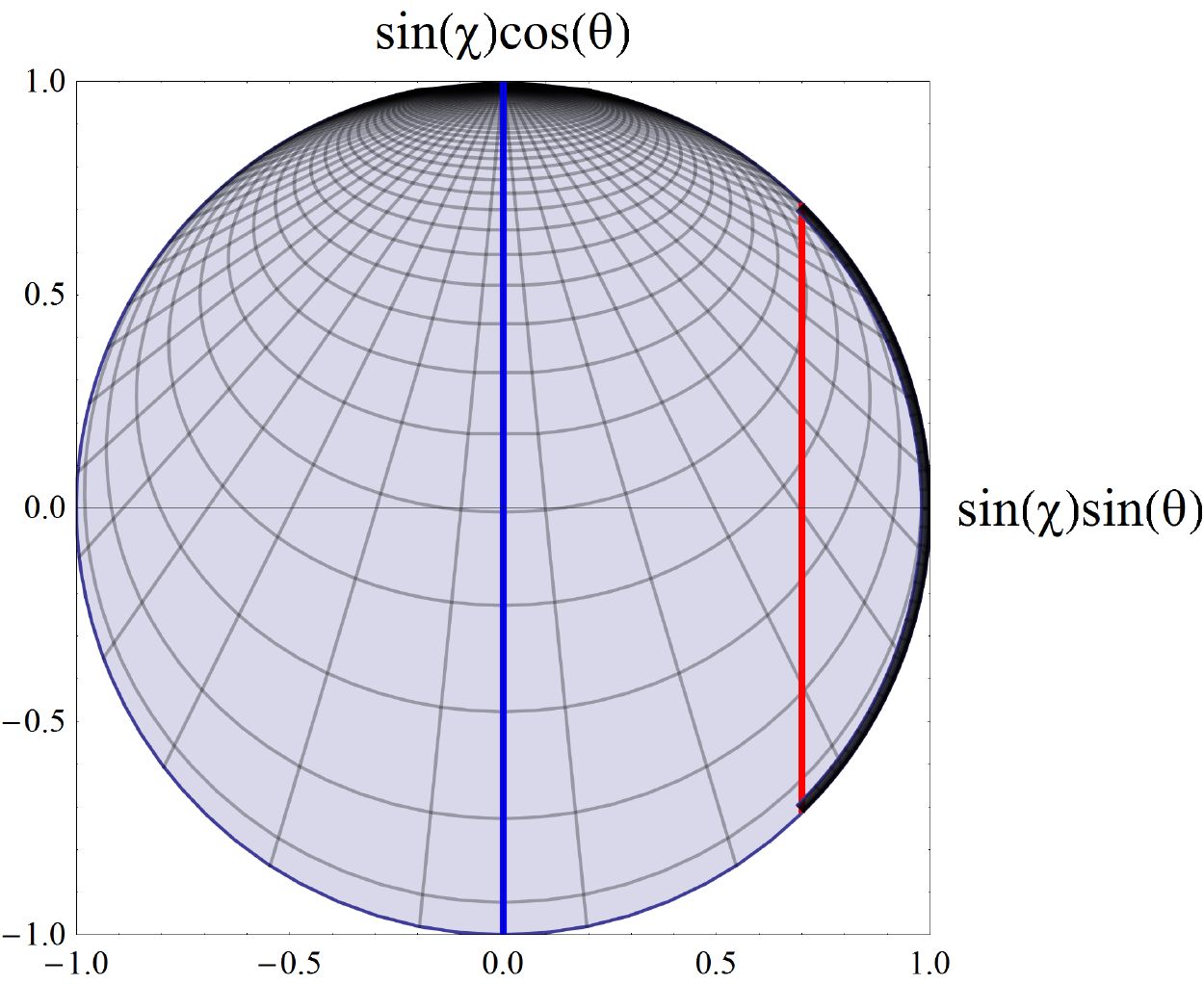}\hspace*{0.02\textwidth}
\includegraphics[width=0.42\textwidth]{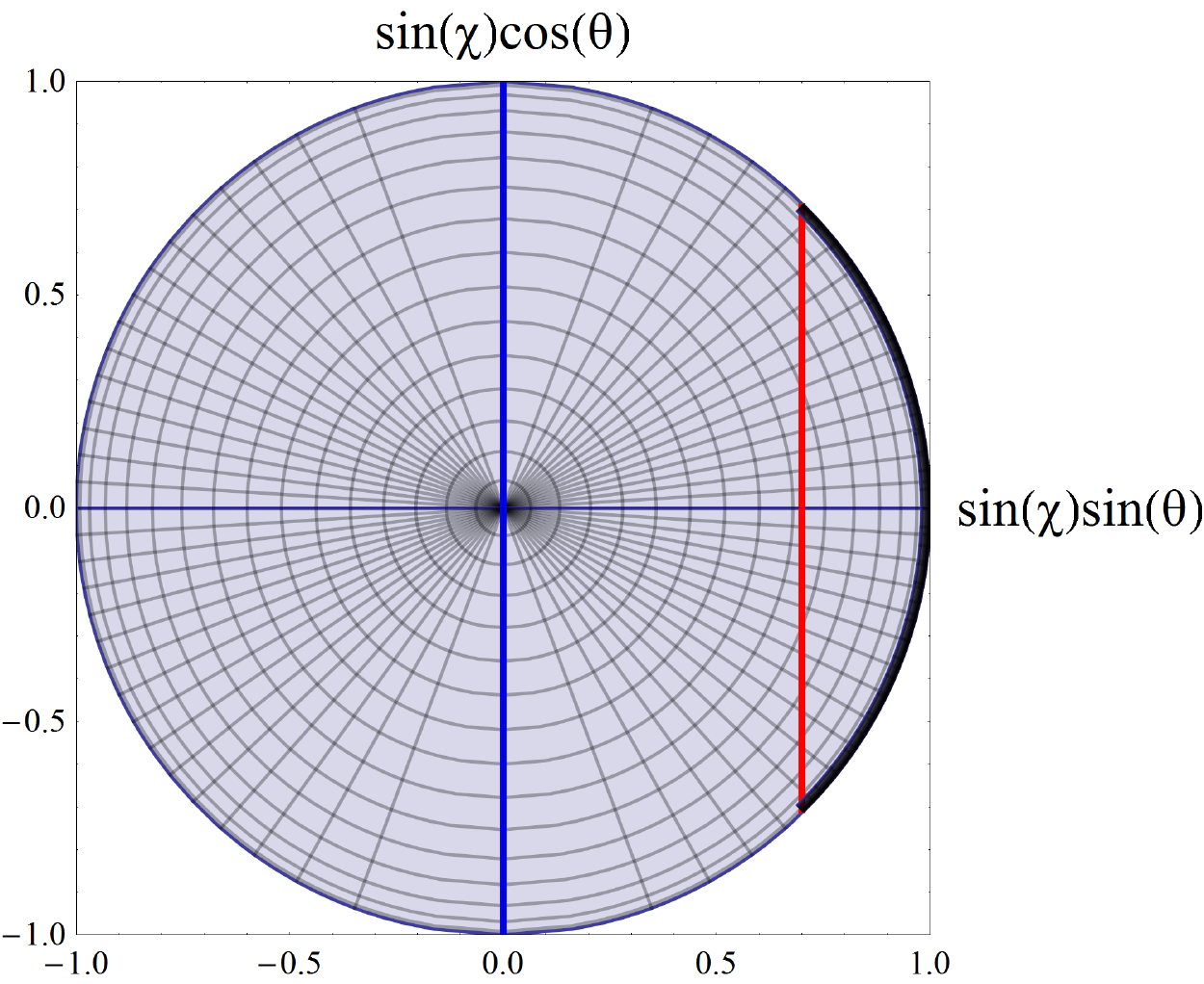}\hspace*{0.02\textwidth}
$$
\caption{Constant-time slice of AdS$_3$ for a Rindler boundary with $a=1$
(left) and a static de Sitter boundary with $H=1$ (right). 
}
\label{cylinder}
\end{figure}

In the left plot of fig. \ref{cylinder} we depict how the slice 
of the AdS$_3$ cylinder with $\eta=0$ is covered by the coordinates 
$y$ and $z$ for $a=1$. 
The two axes correspond to global coordinates. 
The circumference is the AdS$_3$ boundary with $z=0$, which 
is parameterized by the coordinate $y$. The Rindler horizon 
at $y=0$ corresponds to the point $(0,-1)$ in fig. \ref{cylinder}. Positive values
of $y$ cover the right semicircle (R wedge), and negative values the
left semicircle (L wedge). The point $(0,1)$ is approached in the 
limits $y\to \pm \infty$ from right or left. The AdS$_3$ interior is covered by
lines of constant $y$ and variable positive $z$. All these lines converge 
to the point (0,1) for $z\to \infty$. 
We expect to have entanglement between the R and L wedges. The
corresponding entanglement entropy can obtained through holography by 
computing the area of the minimal surface $\gamma_A$ of ref. \cite{ryu,ryu2,ryu3}.
This is depicted by the blue line in this case, which acts as a bulk horizon.
The Rindler horizon can be viewed as the holographic image of the bulk horizon.

Let us consider a strip with width $l$ in the $y$-direction and very large extent in
the remaining spacelike directions.
The minimal
surface extends into the bulk up to \linebreak
$z_*=\Gamma\left(\frac{1}{2d}\right)/\left(2\sqrt{\pi}\,
\Gamma\left(\frac{d+1}{2d}\right)\right)\,l.$ 
In global coordinates this surface corresponds to a straight line through 
the bulk, as depicted by the red line in fig. \ref{cylinder}. 
The entanglement entropy can be computed as 
\be
S_A=
\frac{2R(R^{d-1}L^{d-1})}{4G_{d+2}}
\left(
\frac{1}{(d-1) \epsilon^{d-1}}
+\frac{\sqrt{\pi}}{2d}\frac{\Gamma\left(\frac{1-d}{2d}\right)}{
\Gamma\left(\frac{1}{2d}\right)}
\frac{1}{z_*^{d-1}}
\right).
\label{Rindlerentropy} \ee
A cutoff $\epsilon$ has been imposed on $z$ as the surface
approaches the boundary.
For $d=1$, one must substitute $1/((d-1)\epsilon^{d-1})$ with $\log(1/ \epsilon$). 
Here $L$ is the large length of the
directions perpendicular to the strip, so that $R^{d-1}L^{d-1}$ is the 
corresponding volume.

We are interested in the limit in which the width $l$ of the strip covers the whole
R wedge. In this case the entanglement occurs between the R and L wedges. 
For $l\to \infty$ we have $z_*\to \infty$ and the second  
term in the parenthesis in eq. (\ref{Rindlerentropy}) vanishes. 
In order to assign a physical meaning to the first term, 
we can define the effective
Newton's constant for the boundary theory as in \cite{hawking}:
\be
G_{d+1}=(d-1) \epsilon^{d-1}\frac{G_{d+2}  }{R},
\label{Geff} \ee
with $(d-1)\epsilon^{d-1}$ replaced by 1/$\log(1/ \epsilon$) for $d=1$.
This definition can be justified in the context of the Randall-Sundrum (RS) model \cite{rs,rs2}, which employs only the part of the AdS space with $z>\epsilon$.
The effective low-energy theory includes dynamical gravity with a Newton's
constant given by eq. (\ref{Geff}).
In the limit $\epsilon\to 0$, the constant vanishes and gravity becomes non-dynamical.
This demonstrates the difficulty in computing the gravitational entropy in 
the context of the AdS/CFT coorespondence. The resolution we suggest 
is to keep the cutoff nonzero and absorb it in the definition of the effective
Newton's constant. 
Trading $\epsilon$ for $G_{d+1}$ in the expression for the entropy results in
a meaningful expression.

Substituting eq. (\ref{Geff}) in eq. (\ref{Rindlerentropy})
gives an entanglement
entropy 
which is bigger by a factor of 2 than the known gravitational entropy \cite{laflamme}. 
The reason can be traced to the way the limit is taken in order to cover the whole
R wedge. We start from a strip in the $y$-direction extending 
between two points $y_1$ and $y_2$, and then take the limits $y_1\to 0$ and 
$y_2\to \infty$. The first limit leads to the location of the Rindler horizon.
However, any finite value of $y_2$ excludes an infinite domain corresponding to
$y>y_2$. As a result, the strip is entangled not only with the (infinite) L wedge, but also with the (infinite) domain $y>y_2$. 
The two contributions are expected to be equal because the space is essentially flat. Obtaining
the entropy corresponding to the entanglement with the L wedge only can 
be obtained by dividing the result with a factor of 2. 
The final result for
the Rindler entropy is 
\be
S_R=\frac{R^{d-1}L^{d-1}}{4G_{d+1}},
\label{entropyRindler} \ee
in agreement with \cite{laflamme}.
It is also illuminating to observe that the bulk
horizon depicted as a blue line in fig. \ref{cylinder} approaches the boundary
at two points. The point $(0,-1)$ is the true Rindler horizon. However, the 
point $(0,1)$ does not belong to the boundary Rindler space, but corresponds 
only to the
limits $y\to\pm \infty$. The contribution to the area of the entangling surface
from its vicinity should not be taken into account, thus justifying the
division by 2.

The second case we consider is that of a boundary  
static de Sitter (dS) space:
\begin{equation}
ds^2_{d+2}
= \frac{R^2}{z^2} \left[ dz^2 +\left(1-\frac{1}{4}H^2 z^2 \right)^2 \left(
- (1-H^2\rho^2) dt^2 
+  \frac{d\rho^2}{1-H^2\rho^2}+\rho^2 \, d\Omega^2_{d-1} \right) \right].
\label{dS} \end{equation}
For $d>1$, the range $0\leq \rho \leq 1/H$ covers one static patch. There are two 
such patches in 
the global geometry, which start from the 
the ``North" or ``South pole" at $\rho=0$ and are joined at the surface with 
$\rho=1/H$. 
For $d=1$, $\rho$ can also take negative value and each static
patch is covered by $-1/H \leq \rho \leq 1/H$. 
In the right plot of fig. \ref{cylinder} we depict how the 
slice of the AdS$_3$ cylinder with $t=0$ is covered by the coordinates
$\rho$ and $z$ for $H=1$. 
The circumference is again the AdS$_3$ boundary with $z=0$, which 
is parameterized by the coordinate $\rho$. There are two horizons:
one at $\rho=-1$, corresponding to the point $(0,-1)$, and 
one at $\rho=1$, corresponding to the point $(0,1)$ on the boundary. 
The AdS$_3$ interior is covered by
lines of constant $\rho$ and variable positive $z$. All these lines converge 
to the point (0,0) at the center for $z\to \infty$. 
In the context of the global geometry, we expect to have entanglement between 
the two static patches. The
corresponding entanglement entropy can be obtained through holography by 
computing the area of the minimal surface $\gamma_A$ of ref. \cite{ryu,ryu2,ryu3},
depicted by the blue line. 
This line acts as bulk horizon. 
The difference with the Rindler case we discussed before is that
the endpoints of the minimal surface 
are points of the boundary dS space, they are actually the horizons. 
This means that there is no need to divide by a factor of 2 in this case.
For $d>1$ the $d$-dimensional minimal surface $\gamma_A$ 
ends on an $(d-1)$-dimensional sphere that separates the two
hemispheres of the slice of dS$_{d+1}$ with $t=0$.

The isometries of dS space indicate that the entangling surface is spherical in this
case.
The minimal surface $\gamma_A$ in the bulk 
can be determined by minimizing the integral
\be
{\rm Area}(\gamma_A)=R^d S^{d-1}\int 
d\sigma \frac{\sin^{d-1}(\sigma)}{\sinh^{d}(w)}\sqrt{1+\left(\frac{dw(\sigma)}{d\sigma}\right)^2},
\label{areads2} \ee
where we have defined the parameters $\sigma=\sin^{-1}(H\rho)$, $w=2\tanh^{-1}(Hz/2)$,
and denoted the volume of the ($d-1$)-dimensional unit sphere as $S^{d-1}$.
The above expression is minimized by the function \cite{tetradis2}
\be
w(\sigma)=\cosh^{-1} \left( 
 \frac{\cos(\sigma)}{\cos(\sigma_0)}\right).
\label{solws} \ee 
For $\sigma_0\to 0$ 
the known expression
$w(\sigma)=\sqrt{\sigma_0^2-\sigma^2}$ \cite{ryu,ryu2,ryu3} for 
$H=0$ is reproduced. For $\sigma_0\to \pi/2$ 
the boundary is approached at the location of the 
horizon  with $dw/d\sigma \to -\infty$.

The integral (\ref{areads2}) is dominated by the region near the boundary. 
Introducing a cutoff at $z=\epsilon$ results in a leading
contribution 
\be
{\rm Area}(\gamma_A)=
R^d S^{d-1}I(\epsilon) =
R^d S^{d-1}\int_{H\epsilon} 
\frac{dw}{\sinh^{d}(w)}.
\label{areaeps} \ee 
For $d\not= 1$ the leading divergent part is 
$I(\epsilon)=1/((d-1)H^{d-1}\epsilon^{d-1})$, while for $d=1$ it is
$\log(1/ (H \epsilon$)).
Using eq. (\ref{Geff}) we obtain the  
leading contribution to the entropy: 
\be
S_{\rm dS}=\frac{{\rm Area}(\gamma_A)}{4G_{d+2}}=
\frac{R^d S^{d-1}}{4G_{d+2}(d-1)H^{d-1}\epsilon^{d-1}}=
\frac{S^{d-1}}{4G_{d+1}}\left(\frac{R}{H}\right)^{d-1}
=\frac{A_H}{4G_{d+1}},
\label{entropydss} \ee
with $A_H$ the area of the horizon. This result
reproduces the gravitational entropy of \cite{gibbons}.
It is valid for $d=1$ as well, 
with $1/((d-1)\epsilon^{d-1})$ replaced by $\log(1/ \epsilon$)
and $S^0=2$, because the horizons of the global dS$_2$ geometry are 2 points
\cite{hawking}.

The integral $I(\epsilon)$ also contains subleading divergences. 
There is a subleading logarithmic divergence for $d=3$,
no singular subleading terms for $d=2$, while the only divergence for $d=1$
is the leading logarithmic term already included in eq. (\ref{entropydss}).
For $d>3$ we have subleading power-law divergences for odd $d+1$, plus a 
logarithmic one for even $d+1$.
We focus on four dimensions, in which the 
dS entropy takes the form
\be
S_{\rm dS}=\frac{A_H}{4G_4}\left(
1+H^2 \epsilon^2 \log H \epsilon \right).
\label{holentropydS} \ee
The logarithmic 
dependence on the cutoff hints at a connection with the conformal anomaly of
the dual theory, which results from higher curvature terms in the effective 
theory. The effective action can be deduced from 
known results for the on-shell action in holographic renormalization
\cite{skenderis,skenderis2,skenderis3}. In our approach the divergences are not removed through 
the introduction of counterterms, but are absorbed in the effective couplings.
This means that the relevant quantity for our purposes is the regulated form of the effective action.
Using the results of \cite{skenderis,skenderis2,skenderis3}, we obtain the leading terms \cite{tetradis2}
\be
S=\frac{R^3}{16\pi G_5}\int d^4x\, \sqrt{-\gamma}
\left[ \frac{6}{\epsilon^4}+\frac{1}{2\epsilon^2}{\cal R}
-\frac{1}{4}\log\epsilon 
\left( {\cal R}_{ij}{\cal R}^{ij}-\frac{1}{3}{\cal R}^2
\right) \right].
\label{effaction} \ee
The first term corresponds to a cosmological constant. In the RS model \cite{rs,rs2} this
is balanced by the surface tension of the brane at $z=\epsilon$.
The second term is the standard Einstein term if the effective Newton's
constant $G_4$ is defined as in eq. (\ref{Geff}) with $d=3$. 
The third term is responsible for the holographic conformal anomaly.
The action (\ref{effaction}) supports a dS solution.
In order to take into account the
presence of the higher-curvature terms in eq. (\ref{effaction}) one must compute
the Wald entropy \cite{wald,wald2,wald3}. 
The result is in agreement with the singular part of the
correction provided by the holographic calculation (\ref{holentropydS}) \cite{tetradis2}.

For the ${\cal N}=4$ supersymmetric $SU(N)$ gauge 
theory in the large-$N$ limit,
the  effective action can be computed as \cite{bd}
\be
S=-\frac{\beta}{16\pi^2} \Gamma\left(2-\frac{d+1}{2}\right) 
\int d^4x\, \sqrt{-\gamma}
\left( {\cal R}_{ij}{\cal R}^{ij}-\frac{1}{3}{\cal R}^2
\right),
\label{weyleuler} \ee
with $\beta=-N^2/4$.
The divergence of $\Gamma(2-(d+1)/2)$  in dimensional regularization 
in the limit $d+1\to 4$ corresponds to a $\log(1/\epsilon^2)$ divergence 
in our cutoff regularization.
A comparison of the above expression with eq. (\ref{effaction}) 
reproduces the standard AdS/CFT relation $G_5=\pi R^3/(2N^2)$.
The dimensionful UV momentum cutoff for $d=3$ can be expressed as
$\left( \epsilon_N R \right)^{-2}=2G_5/(R^3 G_4)=8\pi^2 m_{\rm Pl}^2/N^2$, with $m_{\rm Pl}^2=1/(8\pi G_4)$.
Now eq. (\ref{holentropydS}) for $d=3$ can be cast in the form 
\be
S_{\rm dS}
=\frac{A_H}{4G_4}+N^2 \log(H\epsilon_N)
=\frac{A_H}{4G_4}+N^2 \log \left( \frac{N}{\sqrt{8}\pi}\frac{H/R}{m_{\rm Pl}}
\right),
\label{holentropydSdual} \ee
where $H/R$ is the physical Hubble scale.
This expression is completely analogous to the black-hole result \cite{sen}, 
with the horizon size parameter measured in units 
of the UV cutoff. It is also in agreement with the calculation of the
logarithmic part of the holographic entanglement entropy in \cite{chm}.

The calculation of the entropy associated with 
nontrivial gravitational backgrounds through holography faces two difficulties:
\begin{itemize}
\item
The boundary metric in the context of AdS/CFT is not dynamical, a feature
that is equivalent to $m_{\rm Pl} \to \infty$. 
\item
The entanglement entropy has a strong dependence on the UV cutoff of the
theory, which makes its identification with the gravitational entropy
problematic.
\end{itemize}
We showed that these difficulties can be resolved if the UV cutoff
dependence is absorbed in the definition of  $m_{\rm Pl}$. The conceptual
framework is provided by the Randall-Sundrum model \cite{rs,rs2}, or, alternatively,
by the regulated form of the effective action in holographic 
renormalization \cite{skenderis,skenderis2,skenderis3}.
Our derivation of the dS entropy 
is consistent with the expectation that the entropy associated
with gravitational horizons can be understood as 
entanglement entropy if Newton's constant
is induced by quantum fluctuations of matter fields \cite{jacobson,jacobson2}.
In the context of the AdS/CFT correspondence
the bulk degrees of freedom correspond to the matter fields of the 
dual theory. 
The boundary Einstein action arises through the integration
of these bulk degrees of freedom up to the UV cutoff. 

Our approach is in contrast with the usual interpretation of the leading
contribution to the entanglement entropy as 
an unphysical UV-dependent quantity of little interest. We have reached the
opposite conclusion:
The leading contribution to the entropy has a universal
form that depends only on the horizon area because the same degrees
of freedom contribute to the entropy and Newton's constant. Also, the
detailed nature of the 
UV cutoff does not affect the leading contribution. 
The particular features of the underlying theory, such as the number of 
degrees of freedom  
become apparent at the level of the subleading corrections to
the entropy:
the coefficient of the logarithmic correction is determined by the
central charge of the theory.

\section*{Acknowledgments}

This research work was supported by the Hellenic Foundation for Research and Innovation (H.F.R.I.) under the First Call for H.F.R.I. Research Projects to support Faculty members and Researchers (Project Number: 824).




\end{document}